\documentclass[reprint,prl]{revtex4-2}

\usepackage{comment}
\usepackage{xcolor}
\usepackage{graphicx}
\usepackage{amsmath}

\definecolor{updateblack}{rgb}{0.0, 0.0, 0.0}
\newcommand{\upd}[1]{{\color{updateblack}{#1}}}

\begin{document}

\title{First-Principles Approach for Coupled Quantum Dynamics of Electrons and Protons in Heterogeneous Systems}

\author{Jianhang Xu}
\affiliation{Department of Chemistry, University of North Carolina at Chapel Hill}
\author{Ruiyi Zhou}
\affiliation{Department of Chemistry,  University of North Carolina at Chapel Hill}
\author{Volker Blum}
\affiliation{Thomas Lord Department of Mechanical Engineering and Materials Science, Duke University}
\affiliation{Department of Chemistry,  Duke University}
\author{Tao E. Li}
\affiliation{Department of Chemistry, Yale University}
\author{Sharon Hammes-Schiffer}
\email{sharon.hammes-schiffer@yale.edu}
\affiliation{Department of Chemistry, Yale University}
\author{Yosuke Kanai}
\email{ykanai@unc.edu}
\affiliation{Department of Chemistry,  University of North Carolina at Chapel Hill} \affiliation{Department of Physics and Astronomy, University of North Carolina at Chapel Hill}

\date{\today}

\begin{abstract}
The coupled quantum dynamics of electrons and protons is ubiquitous in many dynamical processes involving light-matter interaction, such as solar energy conversion in chemical systems and photosynthesis. 
A first-principles description of such nuclear-electronic quantum dynamics requires not only the time-dependent treatment of nonequilibrium electron dynamics but also that of quantum protons. 
Quantum mechanical correlation between electrons and protons adds further complexity to such coupled dynamics. 
Here we extend real-time nuclear-electronic orbital time-dependent density functional theory (RT-NEO-TDDFT) to periodic systems and perform first-principles simulations of coupled quantum dynamics of electrons and protons in complex heterogeneous systems. 
The process studied is electronically excited state intramolecular proton transfer of o-hydroxybenzaldehyde in water and at a silicon (111) semiconductor-molecule interface. 
These simulations illustrate how environments such as hydrogen-bonding water molecules and an extended material surface impact the dynamical process on the atomistic level. 
Depending on how the molecule is chemisorbed on the surface, excited state electron transfer from the molecule to the semiconductor surface can inhibit ultrafast proton transfer within the molecule. 
This work elucidates how heterogeneous environments influence the balance between the quantum mechanical proton transfer and excited electron dynamics. 
The periodic RT-NEO-TDDFT approach is applicable to a wide range of other photoinduced heterogeneous processes.

\end{abstract}

\maketitle

Quantum dynamics of electrons and protons are essential for various dynamical processes involving light-matter interaction. 
The ability to investigate coupled dynamics of excited electrons and proton transfer from first principles is particularly important in areas of solar energy conversion research. 
Excited-state intramolecular proton transfer (ESIPT) serves as a prototypical example of such a quantum mechanical process and holds significance in numerous biological and chemical systems.\cite{chen2018amino,sedgwick2018excited,jankowska2021modern,joshi2021excited}
Although ESIPT has been the topic of many theoretical studies,\cite{sobolewski1999ab,scheiner2000theoretical,ma2017effect,somasundaram2018structural}
the impact of heterogeneous environments, such as a material surface with interfacial solvating water molecules, on this dynamical process, has not been explored extensively. 
The proton transfer and excited electron dynamics are quantum mechanically coupled, 
and the environment is likely to influence the delicate balance of these coupled quantum dynamics in a complicated manner. 
In particular, excited electrons are inherently dynamic in condensed matter systems due to the dense manifold in the electronic excitation spectrum. 
Although ESIPT typically occurs within a well-defined electronic excited state in an isolated molecular system, 
such proton transfer becomes coupled to the non-equilibrium dynamics of excited electrons when the molecule is situated in a heterogeneous environment. 

Nonequilibrium dynamics of electrons in response to external stimuli can be studied by simulating the time evolution of the quantum state with a time-dependent perturbation in the system’s Hamiltonian.\cite{goings2018real,shepard2021simulating}
In the last few decades, the real-time propagation approach to time-dependent density functional theory (RT-TDDFT)\cite{yabana1996time} has gained popularity as a particularly practical computational methodology for investigating nonlinear electronic responses in complex matter because of the appealing balance between accuracy and efficiency. 
RT-TDDFT simulations have been increasingly employed to address various scientific questions associated with nonequilibrium electronic dynamical phenomena\cite{li2007ab,li2020real}, including those of condensed matter systems\cite{tateyama2006real,yao2019k,noda2019salmon,kononov2022electron,shepard2023electronic}. 

\upd{At the same time, the inclusion of nuclear quantum dynamical effects is important for simulating processes such as excited state proton transfer.
A wide range of approaches have been developed for describing nuclear quantum effects, such as path integral formulations,\cite{berne1986simulation,craig2005chemical,herrero2014path,markland2018nuclear,ananth2022path} exact factorization,\cite{abedi2010exact,li2022energy,villaseco2022exact} and multiconfigurational time-dependent Hartree quantum dynamics\cite{beck2000multiconfiguration,worth2008using}. \citeauthor{zhao2023ring} \cite{zhao2023ring} recently combined a RT-TDDFT approach with ring-polymer dynamics based on the path-integral formulation and showed the importance of proton quantization in a water dimer. 
In addition to these approaches, the nuclear electronic orbital (NEO) method developed by Hammes-Schiffer and co-workers has been shown to be effective for treating specified nuclei quantum mechanically on the same level as electrons within various electronic structure methods.\cite{webb2002multiconfigurational,chakraborty2008development,hammes2021nuclear,pavošević2020multicomponent}
}
The NEO-DFT method is formally based on the general  multicomponent DFT formalism.\cite{capitani1982non,kreibich2001multicomponent,pak2007density,chakraborty2009properties} 
In recent work, we extended the NEO method for periodic electronic structure calculations so that condensed matter systems can be studied.\cite{xu2022nuclear}
The NEO method has also been combined with RT-TDDFT simulations (RT-NEO-TDDFT) for studying the coupled quantum dynamics of protons and electrons in molecular systems.\cite{zhao2020real,zhao2020nuclear,zhao2021excited,li2022semiclassical,li2023electronic}
Importantly, the NEO formulation does not rely on the usual Born-Oppenheimer approximation between electrons and protons and provides real-time quantum dynamics of protons and electrons on equal footing. 

Herein, we develop the periodic RT-NEO-TDDFT approach, which enables first-principles simulations of the non-Born-Oppenheimer, nuclear-electronic quantum dynamics of extended condensed matter systems.
By employing the Kohn-Sham (KS) ansatz within  multicomponent DFT, the system's dynamics are governed by a set of coupled time-dependent KS equations for electrons and quantum protons,
\begin{align*}
    i\frac{\partial}{\partial t} \psi^e_{i\textbf k}(\textbf r^e, t) =  \left[ -\frac{1}{2m^e}\nabla^2 \right. &+v^e_\text{KS}(\textbf r^e) - v^p_\text{es}(\textbf r^e) \\
     & + \left. \frac{\delta E_\text{epc}[\rho^e, \rho^p]}{\delta \rho^e(\textbf r^e, t)} \right] \psi^e_{i\textbf k}(\textbf r^e, t), \\
    i\frac{\partial}{\partial t} \psi^p_{i}(\textbf r^p, t) = \left[ -\frac{1}{2M^p}\right. \nabla^2 &+ v^p_\text{KS}(\textbf r^p) - v^e_\text{es}(\textbf r^p) \\
    &  + \left. \frac{\delta E_\text{epc}[\rho^e, \rho^p]}{\delta \rho^p(\textbf r^p, t)} \right] \psi^p_{i}(\textbf r^p, t).
\end{align*}
Here, $\psi^e_{i\textbf k}(\textbf r^e, t)$ and $\psi^p_{i}(\textbf r^p,t)$ are the electronic and protonic time-dependent KS orbitals, and $m^e$ and $M^p$ are the electron and proton masses, respectively. 
$v^{e/p}_\text{KS}$ is the standard KS effective potential while $v^{e/p}_\text{es}$ represents the electrostatic potential due to the other types of quantum particles.
$E_\text{epc}$ is the quantum mechanical correlation functional between electrons and protons.\cite{chakraborty2008development,yang2017development,brorsen2017multicomponent}
This electron-proton correlation (EPC) is important for accurately describing several proton properties such as the proton densities and zero-point energy in previous works.\cite{pavošević2020multicomponent,tao2021analytical} 
Technical implementation details of the periodic RT-NEO-TDDFT approach in the all-electron FHI-aims code\cite{blum2009ab,hekele2021all} are available in the Supplemental Material.
Using this newly developed periodic RT-NEO-TDDFT method, we elucidate the competing kinetics between excited electron dynamics and quantum mechanical proton transfer in heterogeneous environments.

\begin{figure*}[t]
    \centering
    \includegraphics[width=0.92\textwidth]{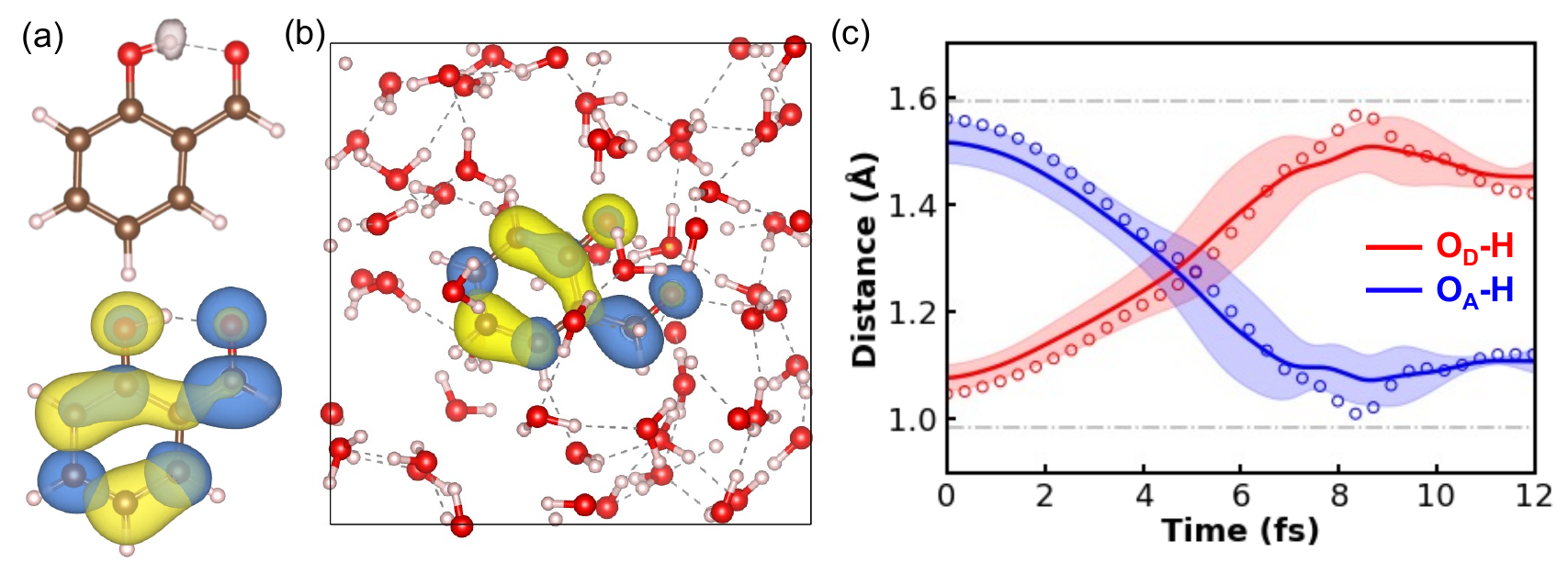}
    \caption{The geometries of (a) the isolated oHBA molecule (with the quantum proton highlighted using a white isosurface)  and (b) a snapshot of the oHBA molecule solvated in liquid water from FPMD simulation. The isosurfaces show the molecular HOMO state (yellow) and the molecular LUMO state (blue).
    (c) Distances between the position expectation value of the transferring proton and the donor (blue) and acceptor (red) oxygen atoms in the oHBA molecule as a function of time. The circles represent the isolated oHBA molecule in vacuum, and the solid lines represent the solvated oHBA molecule in water. The shaded areas along the solid lines indicate the standard deviation for the ensembles using six different snapshots from the FPMD simulation.}
    \label{fig:oHBA-water} 
\end{figure*}

Quantum mechanical proton transfer between two oxygen atoms in the o-hydroxybenzaldehyde (oHBA) molecule is a well-known example of ESIPT. 
In a previous study, \citeauthor{zhao2020real} investigated the transfer dynamics using RT-NEO-TDDFT and observed ultrafast proton transfer.\cite{zhao2020real}
In this work, the initial electronic excited state was prepared by promoting one electron from the molecule's HOMO to its LUMO and, to facilitate ESIPT, the molecular geometry was chosen to be the structurally relaxed geometry in the S1 excited state with the proton constrained to be bonded to its donor \cite{zhao2020real} (See Fig. \ref{fig:oHBA-water}(a)).
All nuclei other than the transferring proton were fixed in the initial work, but the Ehrenfest dynamics of these classical nuclei were also simulated in later work.\cite{zhao2021excited}
These simulations were conducted on an isolated oHBA molecule in vacuum or in dielectric continuum water\cite{wildman2022solvated}.

Herein, we investigate situations where the oHBA molecule is in heterogeneous environments, including the presence of explicit water molecules and adsorption onto a semiconductor surface. 
Such scenarios are commonly encountered in solar-fuel research\cite{dempsey2021vision} where photo-/electro-active molecules are adsorbed on semiconductor surfaces. 
Nevertheless, our current understanding of how various environmental effects, such as hydrogen bonding from solvating water molecules or electrostatic fields from the semiconductor surface, impact the coupled quantum dynamics between protons and electrons is limited.
\upd{Our RT-NEO-TDDFT simulations employ the PBE\cite{perdew1996generalized} generalized-gradient approximation, together with the adiabatic approximation\cite{lacombe2023non}}, to the exchange-correlation (XC) functional. The Tier 2 numeric atom-centered orbitals (NAO)\cite{blum2009ab} basis set is employed for electrons. The protons were modeled with the Hartree-Fock approximation and an sp Gaussian-type basis with the fixed proton basis function center method, following Ref. \cite{zhao2020real}.
The epc17-2 electron-proton correlation functional\cite{yang2017development,brorsen2017multicomponent}, which was developed based on an extension of the Colle-Salvetti formulation, was used for the electron-proton correlation. 
\upd{Additional computational details are provided in the Supplemental Material, including the comparison to conventional RT-TDDFT simulations to illustrate the effect of the quantum proton dynamics on the excited electron dynamics.   
}

Aqueous solutions are typical condensed matter environments for many molecular photochemical processes. 
Molecules solvated in aqueous solutions are particularly ubiquitous in solar-fuel conversion processes involving CO$_2$ reduction and H$_2$O oxidation. 
To obtain equilibrated configurations of oHBA in water, we performed first-principles molecular dynamics (FPMD) simulations with the oHBA molecule in a cubic simulation cell using periodic boundary conditions (PBC) with 57 water molecules in each simulation cell at room temperature (300K). 
For these FPMD simulations, we used the SCAN meta generalized gradient approximation\cite{sun2015strongly} XC functional, as it has been shown to provide a reasonably accurate structure of water \cite{xu2019first}, and the geometry of the oHBA molecule was held fixed.
We then performed RT-NEO-TDDFT simulations on six randomly selected snapshots from the FPMD trajectory (see Fig. \ref{fig:oHBA-water}(b)). 
\upd{The transferring proton in the oHBA molecule and the nearest protons of the surrounding H$_2$O molecules that are hydrogen-bonded to the two oxygen atoms of the oHBA molecule were quantized.} 

The RT-NEO-TDDFT simulations were initiated by exciting an electron from the HOMO to the LUMO of the oHBA molecule, which are also the valence band maximum (VBM) and conduction band minimum (CBM) states of the entire system in this case. 
Fig. \ref{fig:oHBA-water}(c) shows the time evolution of the proton transfer coordinates, which are defined by the distances from the position expectation value of the transferring quantum proton to the donor oxygen atom (H-O$_{\text{D}}$) and to the acceptor oxygen atom (H-O$_{\text{A}}$), in comparison to the corresponding values from the simulation of the isolated oHBA molecule in vacuum.
The shaded area indicates the variations among the RT-NEO-TDDFT simulations using different snapshots from FPMD as initial conditions, and the solid lines represent the average values. 
Initially, the proton is bonded to the donor oxygen atom, O$_\text{D}$, with the bond length of 1.05 $\text{\AA}$ while the distance to the acceptor oxygen atom, O$_\text{A}$, is 1.56 $\text{\AA}$. 
When the oHBA molecule is isolated in vacuum, the quantum proton transfers in this electronically excited state on the femtosecond timescale, as previously reported.\cite{zhao2020real}
Although the solvating water molecules induce relatively large fluctuations among the six trajectories, ESIPT consistently takes place on a similar time scale compared to the isolated vacuum case. 
\upd{At the same time, the surrounding water molecules influence the details of the quantum proton transfer, including the transfer timescale and ${\text{O}_\text{A}\text{-H/}\text{O}_\text{D}\text{-H}}$ distance oscillations, as indicated by the standard deviation in Fig. \ref{fig:oHBA-water}(c) (see also Supplementary Material).}

Examining the time-dependent probability amplitudes of the excited electron and the hole in terms of the Kohn-Sham (KS) eigenstates (see Supplemental Material) showed that both the excited electron and the hole tend to remain in the HOMO and LUMO of the molecule without acquiring non-molecular characters during the ESIPT process.
In other words, no charge transfer takes place between the oHBA molecule and its surrounding water molecules.
This is not surprising because the molecular HOMO and LUMO are spatially well-localized on the molecule and energetically well-separated from the rest of the valence band and conduction band states (see Supplemental Material). 
In addition, no quantum proton transfer between water molecules and the oHBA molecule was observed. 
Therefore, to a large extent, the ESIPT remains similar to the process in vacuum or dielectric continuum water, even when the molecule is solvated in explicit water. 


\begin{figure}[t]
    \centering
    \includegraphics[width=0.48\textwidth]{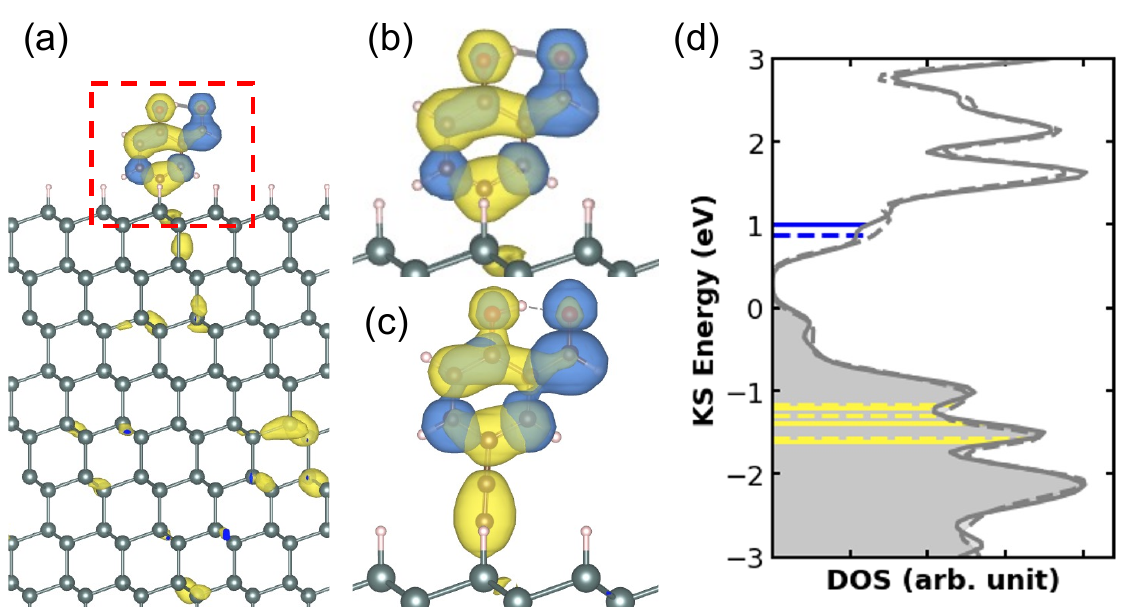}
    \caption{
    (a) Geometry of the oHBA molecule directly chemisorbed on a hydrogen-terminated Si(111) surface. Isosurfaces for one of the three hybridized HOMO (hyb-HOMO) states (yellow) and the hybridized LUMO (hyb-LUMO) state (blue) are also shown.
    Molecular structures of the oHBA molecule with a hyb-HOMO state (yellow) and hyb-LUMO state (blue) are shown for (b) the direct attachment and (c) the attachment with a -C$\equiv$C- linker group in between. (d) The density of electronic states for the direct attachment (solid line) and with the linker group (dashed line). The Fermi energy, \upd{defined here as the valence band maximum (VBM) of the equilibrium ground state,} is at 0 eV, and the hyb-HOMO states and hyb-LUMO state are highlighted in yellow and blue, respectively.}
    
    \label{fig:oHBA-Si-stat}
\end{figure}

Next, we consider the oHBA molecule chemisorbed on a hydrogen-terminated Si(111) surface as an example that is often seen in solar-fuel research.\cite{dempsey2021vision}
The semiconductor surface is modeled using a slab model with eight silicon bilayers (32 Si atoms per bilayer and a total of $\sim$3700 electrons). 
$\Gamma$-point only integration of the Brillouin zone sampling was employed here.
\upd{The transferring proton in the oHBA molecule was quantized.}
As shown in Fig. \ref{fig:oHBA-Si-stat}(b) and (c), two different attachments to the surface are considered:
one with direct chemisorption and the other with a -C$\equiv$C- linker group between the oHBA molecule and the surface. 
When chemisorbed onto the semiconductor surface, the HOMO and LUMO states of the oHBA molecule hybridize with the electronic states of the semiconductor surface.
By using the Mulliken population analysis, we identified three valence band states that derive from the hybridization of the molecule's HOMO and one conduction band state that derives from the molecule's LUMO, as shown in Fig. \ref{fig:oHBA-Si-stat}. Fig. \ref{fig:oHBA-Si-stat}(d) shows that these hybridized molecular states are situated energetically within the manifold of semiconductor electronic states; 
these hybridized HOMO and LUMO states from the oHBA are energetically lower than the VBM and higher than the CBM, respectively. 
We performed several RT-NEO-TDDFT simulations with the initial particle-hole excitation corresponding to an electron excited from one of the HOMO-hybridized (hyb-HOMO) states into the LUMO-hybridized (hyb-LUMO) state. 
Note that the hyb-HOMO states have greater contributions from the surface than does the hyb-LUMO state (Fig. \ref{fig:oHBA-Si-stat}(a)).


\begin{figure}[t]
    \centering
    \includegraphics[width=0.40\textwidth]{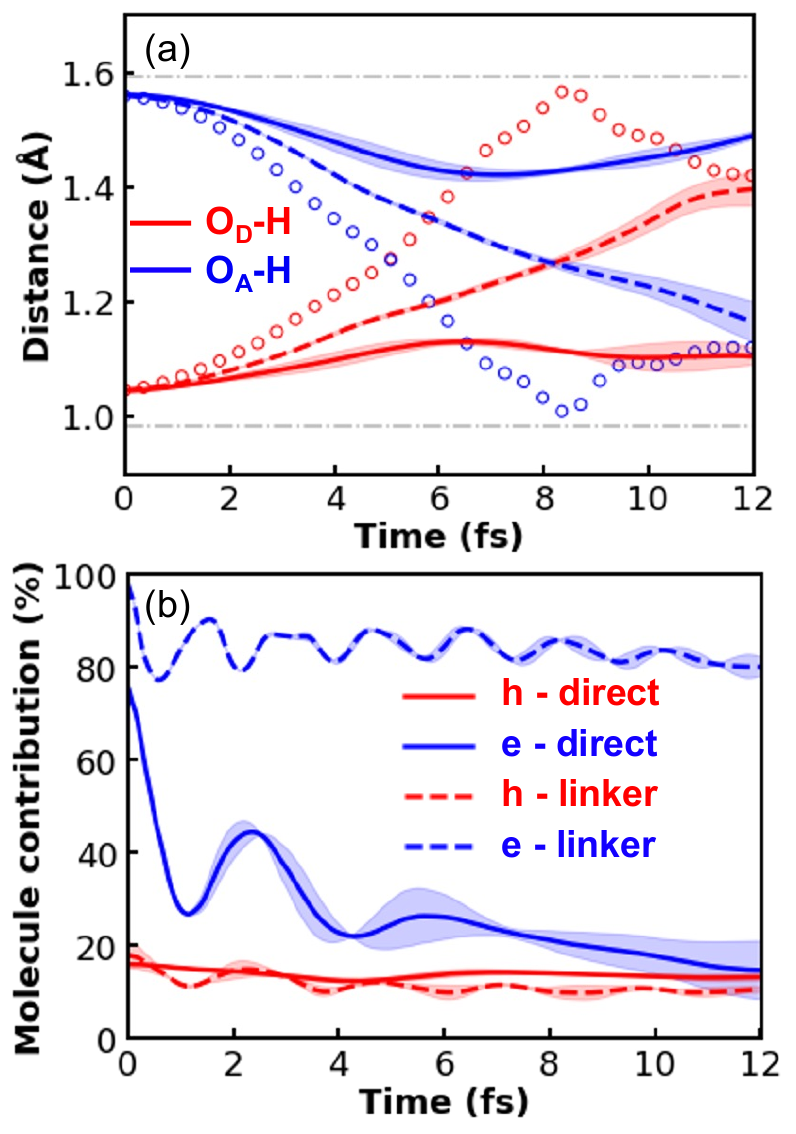}
    \caption{(a) Distances from the position expectation value of the transferring proton to the donor oxygen atom (O$_\text{D}$) (red) and the acceptor oxygen atom (O$_\text{A}$) (blue) as a function of time for the isolated oHBA in vacuum (circles) and the oHBA on the Si surface with the direct attachment (solid line) and with the linker group (dashed line). (b) The percentage of the hole (red) and the excited electron (blue) spatially localized on the oHBA molecule based on Mulliken population analysis with the direct attachment (solid line) and with the linker group (dashed line). The shaded areas along the lines indicate the standard deviation for the initial particle-hole excitation from the three different hyb-HOMO states.}
    \label{fig:oHBA-Si-rt}
\end{figure}

Fig. \ref{fig:oHBA-Si-rt}(a) depicts the proton transfer in terms of the position operator expectation value for the RT-NEO-TDDFT simulations performed with the Si surface. 
The results are compared to those for the isolated molecule in vacuum (shown with open circles). 
The solid and dashed lines correspond to the direct attachment of the molecule onto the surface and attachment through the linker, respectively. 
The shaded areas represent the small variations observed with the excitation of an electron from different hyb-HOMO states (shown in Fig. \ref{fig:oHBA-Si-stat}(d)).
The dynamics are minimally influenced by the specific hyb-HOMO state from which the electron is excited. 
Our RT-NEO-TDDFT simulations show that the ultrafast ESIPT does not occur when the molecule is directly chemisorbed to the surface. 
In contrast, the ESIPT still occurs when the molecule is attached to the surface via the linker group.
It is commonly assumed that the molecule remains in a particular electronically excited state during such an ESIPT process, as prescribed by the Born-Oppenheimer approximation. 
However, the excited electrons are inherently dynamic, and coupling to a dense manifold of electronic states, as for the Si surface, can significantly influence the ESIPT behavior. 

In order to gain insight into the notable difference when the linker group is present, we quantified the spatiotemporal changes for the excited electron and hole using the Mulliken population analysis \cite{mulliken1955electronic}.  
Fig. \ref{fig:oHBA-Si-rt}(b) shows the probability amplitudes of the excited electron and hole within the oHBA molecule.
Because of the hybridization with the surface for hyb-HOMO/LUMO states, the excited electron and hole show significant contributions from the surface instead of the molecule. 
The hole, which exhibits a significantly reduced contribution from the molecule, does not show appreciable changes during the dynamics. 
When the molecule is directly chemisorbed on the surface, the excited electron rapidly transfers to the Si surface (Fig. \ref{fig:oHBA-Si-rt}(b)), which apparently competes with the proton transfer process and effectively suppresses the ESIPT. 
However, with the linker group, the excited electron transfer to the surface is significantly slowed down and the excited electron remains mainly within the molecule on the timescale of the ESIPT. 
Thus, the quantum proton transfer can still take place albeit with a slower transfer time as seen in Fig. \ref{fig:oHBA-Si-rt}(a).

\begin{figure}[t]
    \centering
    \includegraphics[width=0.48\textwidth]{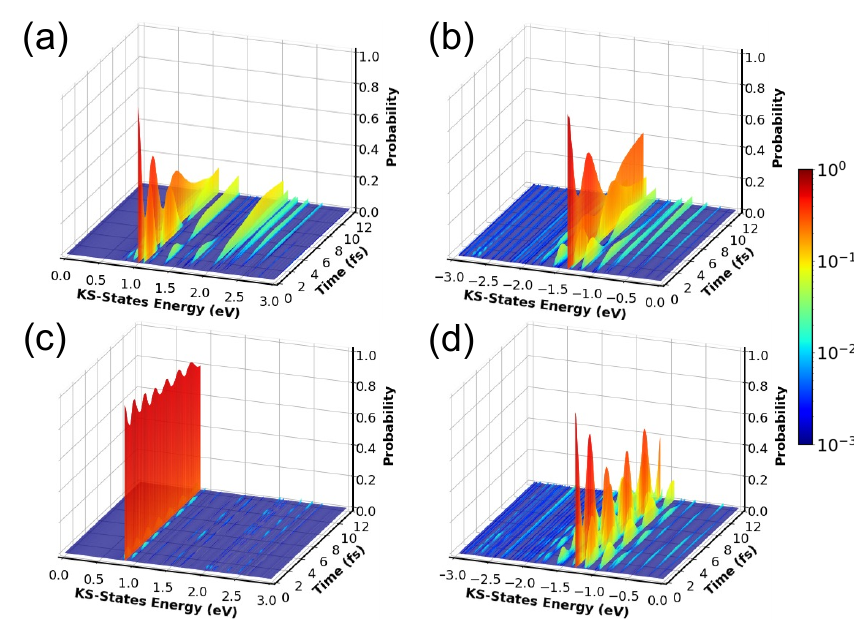}
    \caption{The time-dependent probability amplitudes of the excited electron and hole in terms of the Kohn-Sham eigenstates of the ground-state system with direct attachment (a,b) and with the linker group (c,d). (a,c) and (b,d) are for the excited electron and the hole, respectively.}
    \label{fig:oHBA-Si-3d}
\end{figure}

To gain further insight into the nonequlibirum electron dynamics responsible for the ESIPT, we show the time-dependent probability amplitudes of the hole and the excited electron represented as a superposition of the Kohn-Sham (KS) energy eigenstates from the conduction and valence band states in Figure \ref{fig:oHBA-Si-3d}. 
Fig. \ref{fig:oHBA-Si-3d}(a) and (b) are for the direct chemisorption of the molecule, whereas Fig. \ref{fig:oHBA-Si-3d}(c) and (d) are for the chemisorption with the linker group.  
The figures here show the representative case where an electron is excited from the second hyb-HOMO state into the hyb-LUMO state, and other cases are shown in the Supplemental Material.  

Initially, the excited electron and hole are represented by individual eigenstates, specifically the hyb-LUMO and hyb-HOMO states, respectively. 
As the quantum proton responds to the particle-hole excitation and begins to transfer from the O$_\text{D}$ to the O$_\text{A}$ atom, the excited-state electronic structure changes as well. 
For the case of direct chemisorption, both the excited electron (Fig. \ref{fig:oHBA-Si-3d}(a)) and the hole (Fig. \ref{fig:oHBA-Si-3d}(b)) acquire growing contributions from other eigenstates in time. 
By examining the projected density of states (see Supplemental Material), these eigenstates are determined to be spatially localized on the Si surface for the excited electron. 
The probability amplitude change from the initial hyb-LUMO to other eigenstates is not monotonic but rather is somewhat oscillatory, as expected from the damped oscillatory transfer of the excited electron to the Si surface (Fig. \ref{fig:oHBA-Si-rt}(b)). 
Higher-energy eigenstates show significant amplitudes on this ultrafast timescale, but the excited electron is likely to relax toward the CBM states on a much longer timescale by coupling with phonons.\cite{li2016excited,li2017examining,li2019modeling}
The hole shows a significant oscillatory behavior in the probability amplitudes among the three hyb-HOMO states, as perhaps expected. 
With the presence of the linker group, however, the probability amplitudes show a different behavior. 
For the excited electron (Fig. \ref{fig:oHBA-Si-3d}(c)), the amplitude remains dominated by the initial hyb-LUMO state, as expected from the lack of significant transfer to the Si surface (Fig. \ref{fig:oHBA-Si-rt}(b)).
The hole shows strong periodic amplitude changes, mostly among the three hyb-HOMO states, as seen in Fig. \ref{fig:oHBA-Si-3d}(d).

In summary, we have presented the periodic RT-NEO-TDDFT methodology as a valuable first-principles approach for studying the coupled quantum dynamics of electrons and protons in extended systems beyond the conventional Born-Oppenheimer approximation. 
Through the investigation of ESIPT in the oHBA molecule, we have demonstrated the method's capability to provide mechanistic insights into how this dynamic process is influenced by heterogeneous environments, such as solvating water molecules and semiconductor-molecule interfaces.
Our findings reveal that the ultrafast ESIPT process is highly sensitive to the molecular details of surface adsorption.
Specifically, the presence of a linker group changes the delicate balance between quantum mechanical proton transfer and non-equilibrium dynamics of excited electrons, underscoring the importance of the interplay between electrons and protons.
With the continued development of accurate exchange-correlation approximations and increasingly more powerful computers, we envision the new periodic RT-NEO-TDDFT methodology making important contributions to developing a deeper scientific understanding of coupled quantum dynamics of electrons and protons in complex chemical systems. 
\subsection{ACKNOWLEDGEMENT}
This work is based upon work solely  supported by the Center for Hybrid Approaches in Solar Energy to Liquid Fuels, CHASE, an Energy Innovation Hub funded by the U.S. Department of Energy, Office of Basic Energy Sciences, Office of Science, under award number DE-SC0021173. This research used resources of the National Energy Research Scientific Computing Center, a DOE Office of Science User Facility supported by the Office of Science of the U.S. Department of Energy under Contract No. DE-AC02-05CH11231 using NERSC award BES-ERCAP0024831.

\nocite{kramer2005geometry}
\nocite{magnus1954exponential}
\nocite{gomez2018propagators}
\nocite{zhu2018self}
\nocite{moler1978nineteen}
\nocite{aquino2005excited}
\nocite{xu2019first}

\end{document}